\documentclass[12pt,preprint]{aastex}
\begin{document}

\title{High cadence observations of a global coronal wave by EUVI/STEREO}

\author{Astrid~M. Veronig}
\affil{Institute of Physics/IGAM, University of Graz,
Universit\"atsplatz 5, A-8010 Graz, Austria; asv@igam.uni-graz.at}

\author{Manuela Temmer}
\affil{Space Research Institute, Austrian Academy of Sciences,
Schmiedlstra{\ss}e 6, A-8042 Graz, Austria; manuela.temmer@uni-graz.at}

\author{Bojan~Vr\v{s}nak}
\affil{Hvar Observatory, Faculty of Geodesy, Ka\v ci\' ceva 26,
HR-10000 Zagreb, Croatia; bvrsnak@geodet.geof.hr}

\begin{abstract}
We report a large-scale coronal wave (so-called ``EIT wave") observed with high 
cadence by EUVI onboard STEREO in association
with the GOES B9.5 flare and double CME event on 19 May 2007.
The EUVI instruments provide us with the unprecedented opportunity to study the {\it dynamics}
of flare/CME associated coronal waves. 
The coronal wave under study reveals deceleration, indicative of a freely propagating MHD wave.
Complementary analysis of the associated flare and erupting filament/CME hint at wave initiation by the
CME expanding flanks, which drive the wave only over a limited distance. The associated flare is
very weak and occurs too late to account for the wave initiation.
\end{abstract}

\section{Introduction}

Large-scale large-amplitude waves and shocks in the solar corona occur in association with flares and coronal mass ejections (CMEs). The existence of flare-related global disturbances has been first inferred from
Moreton waves \citep{moreton60}, which appear as arc-like fronts in chromospheric H$\alpha$ filtergrams, moving away from the ignition site with typical velocities of 500--1000~km/s.
It was soon recognized that Moreton waves could not be propagating in the chromosphere, where no wave mode has such high velocity (e.g.\ sound speed and Alfv\'en speed are only of the order of tens of km/s).
The first interpretation was by \cite{uchida68} that Moreton waves are the surface-track of a coronal fast-mode magnetohydrodynamic (MHD) wave front.

The Extreme-ultraviolet (EUV) Imaging Telescope \citep[EIT;][]{dela95} onboard Solar and Heliospheric Observatory (SOHO) for the first time directly imaged propagating global disturbances in the corona, and these so-called ``EIT waves'' were assumed to be the coronal counterparts of the Moreton waves \citep{thompson98,thompson99}. 
Thereafter, coronal waves were found to be a quite frequent phenomenon, and it became an intense matter of debate whether EIT waves: 
\begin{itemize}
 \item[a)] are really the coronal counterparts of Moreton waves
\citep[e.g.][]{thompson00,klassen00,warmuth01,warmuth04a,eto02,khan02,narukage02,vrsnak02,gilbert04,veronig06};
 \item[b)] are caused by the flare explosive energy release or by the erupting CME
\citep[e.g.][]{warmuth01,warmuth04b,biesecker02,hudson03,zhukov04,cliver05,vrsnak06};
 \item[b)] are waves at all (and, if yes, which type of waves; cf.\ \citeauthor{wills07} \citeyear{wills07}) 
 or rather propagating disturbances related to magnetic field line opening and
restructuring associated with the CME lift-off
\cite[e.g.][]{delannee99,wills99,delannee00,wang00,wu01,warmuth01,chen02,vrsnak02,ballai05,attrill07}.
\end{itemize}
In addition, there might be different types of EIT waves, further complicating this debate. For
detailed discussions we refer to the recent reviews by \cite{chen05,vrsnak05,mann07,warmuth07}.

One important limitation of coronal wave studies so far is the low cadence of the EIT instrument (12--15~min),
which makes it impossible to study wave kinematics beyond a rough velocity estimate.
Observations of large-scale waves in TRACE EUV images are rare due to its limited field of view, but we
note that one such event was observed with high cadence and studied in detail in \cite{wills99}.
The Extreme Ultraviolet Imagers (EUVI) on the recent Solar Terrestrial Relations Observatory (STEREO) spacecraft regularly perform EUV full-disk imaging with a cadence as good as 2.5~min. In this letter, we study for the first time the dynamical evolution of a globally propagating ``EIT" wave in high-cadence EUVI
images.

\section{Data}

EUVI is part of the Sun Earth Connection Coronal and Heliospheric Investigation
\citep[SECCHI;][]{howard08} instrument suite onboard STEREO \citep{kaiser08}.
STEREO consists of two identical spacecraft, which orbit the Sun ahead
\mbox{(STEREO-A)} and behind \mbox{(STEREO-B)} the Earth near the ecliptic plane. EUVI is
observing the chromosphere and low corona in four EUV bandpasses (He~{\sc ii} 304~{\AA},
Fe~{\sc ix} 171~{\AA}, Fe~{\sc xii} 195~{\AA}, Fe~{\sc xv} 284~{\AA}) out to 1.7\,$R_s$ (with $R_s$
the solar radius) with a pixel limited spatial resolution of 1.6$''$/pixel.
During the event under study, the EUVI imaging cadence was 2.5 min in the 171~{\AA}
and 10 min in the 195~{\AA} filter.

The impulsive phase of the associated B9.5/SF flare was fully captured in hard X-rays by the Ramaty High Energy Spectroscopic Imager \citep[RHESSI;][]{lin02}. For the study of associated CMEs, we use data from the STEREO/SECCHI inner coronagraph COR1, which has a field-of-view (FOV) from 1.4 to 4\,$R_s$ 
\citep{howard08}, and from the Large Angle Spectroscopic Coronagraph \citep[LASCO;][]{brueckner95} onboard SOHO.

\section{Results}

\subsection{Kinematics and dynamics of the coronal wave}

The coronal wave under study occurred on 19 May 2007 during $\approx$12:50--13:20~UT, in association with the weak B9.5 flare/CME event in AR 10956 close to Sun center (N01$^{\circ}$,W05$^{\circ}$). The two STEREO spacecraft were
8.6$^\circ$ apart and both observed the wave. We concentrate on STEREO-A observations, since it observed a larger portion of the Western solar hemisphere, into which the wave propagated.
For comparison of STEREO-A images with other space-borne and ground-based data, we rescaled
the image sizes to Earth distance. On 19 May 2007, STEREO-A was at a distance of 0.96~AU from Sun,
and Earth was at 1.01~AU. Thus, we decreased the angular diameter of the STEREO-A images by 5.4\%.

\begin{figure}
\epsscale{0.8}
 \plotone{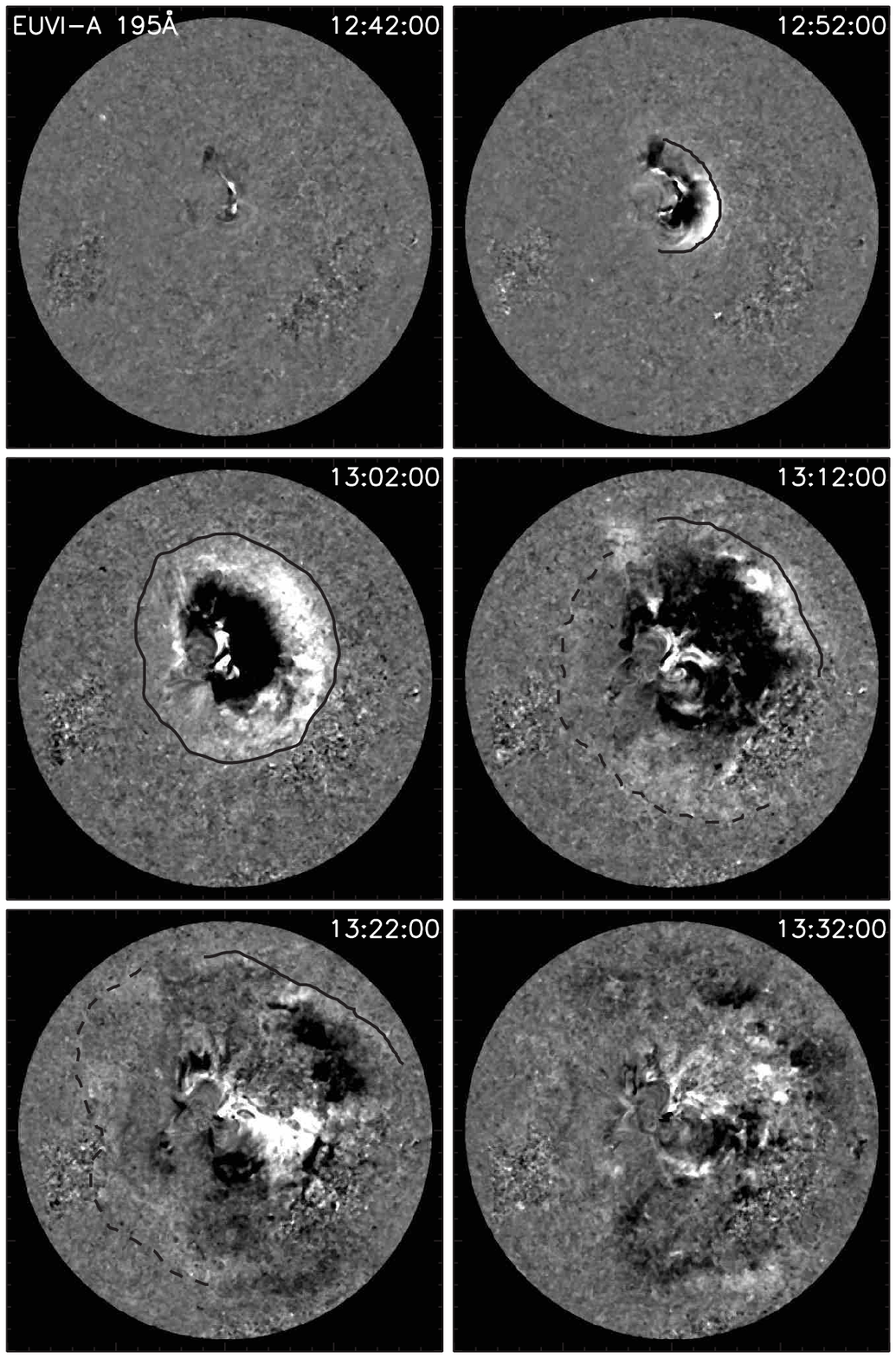}
  \caption{Sequence of median-filtered running ratio images recorded in the EUVI/STEREO-A 195~{\AA} channel
  with a cadence of 10 min.  The identified front edges of the wave
  are indicated with black lines. Only the strongest wave front segments (full lines) are considered in the
  kinematical plot in Fig~\ref{euvi_kinematics}. The plotted FOV is 2000$''\times$2000$''$ around Sun center.
  }
    \label{euvi_195}
\end{figure}

\begin{figure}
\epsscale{0.8}
 \plotone{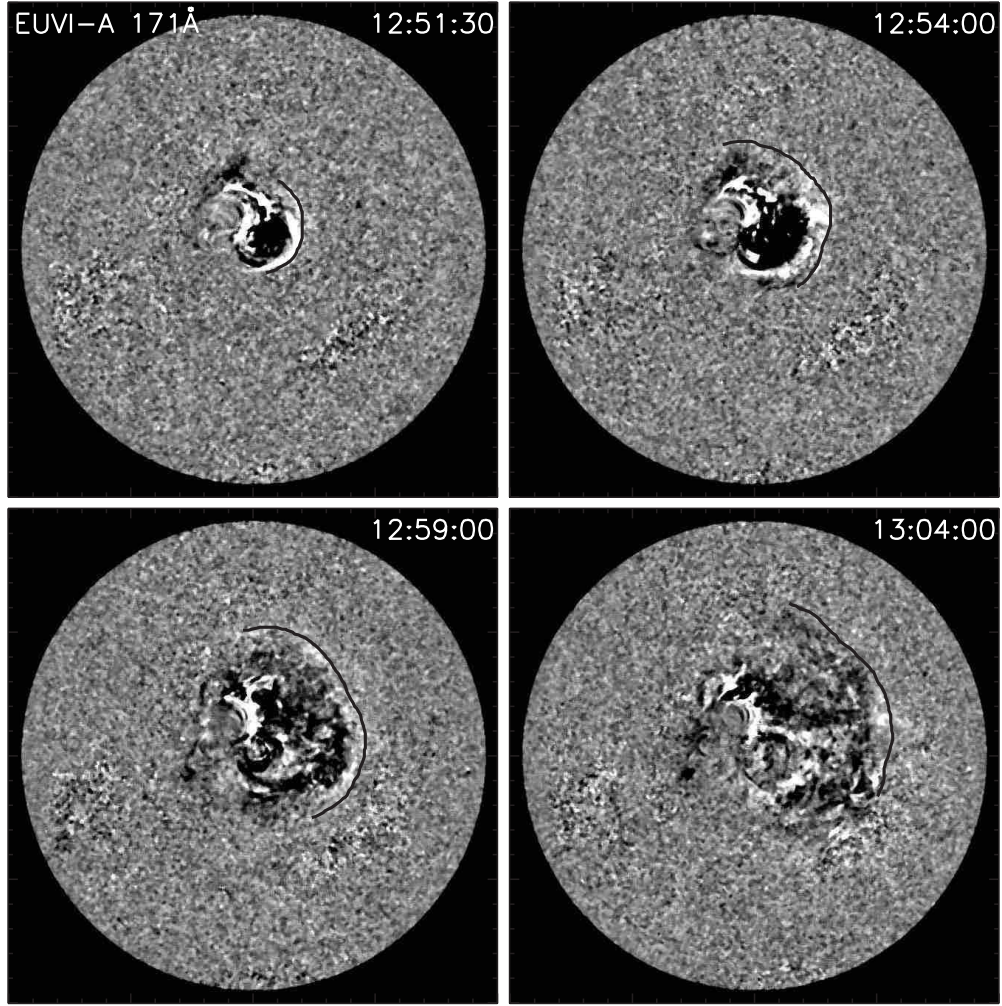}
  \caption{Same as Fig.~\ref{euvi_195} but for the EUVI 171~{\AA} channel. Not all images
  in which the wave can be identified (2.5~min cadence) are shown.}  \label{euvi_171}
\end{figure}

\begin{figure}
\epsscale{1}
 \plotone{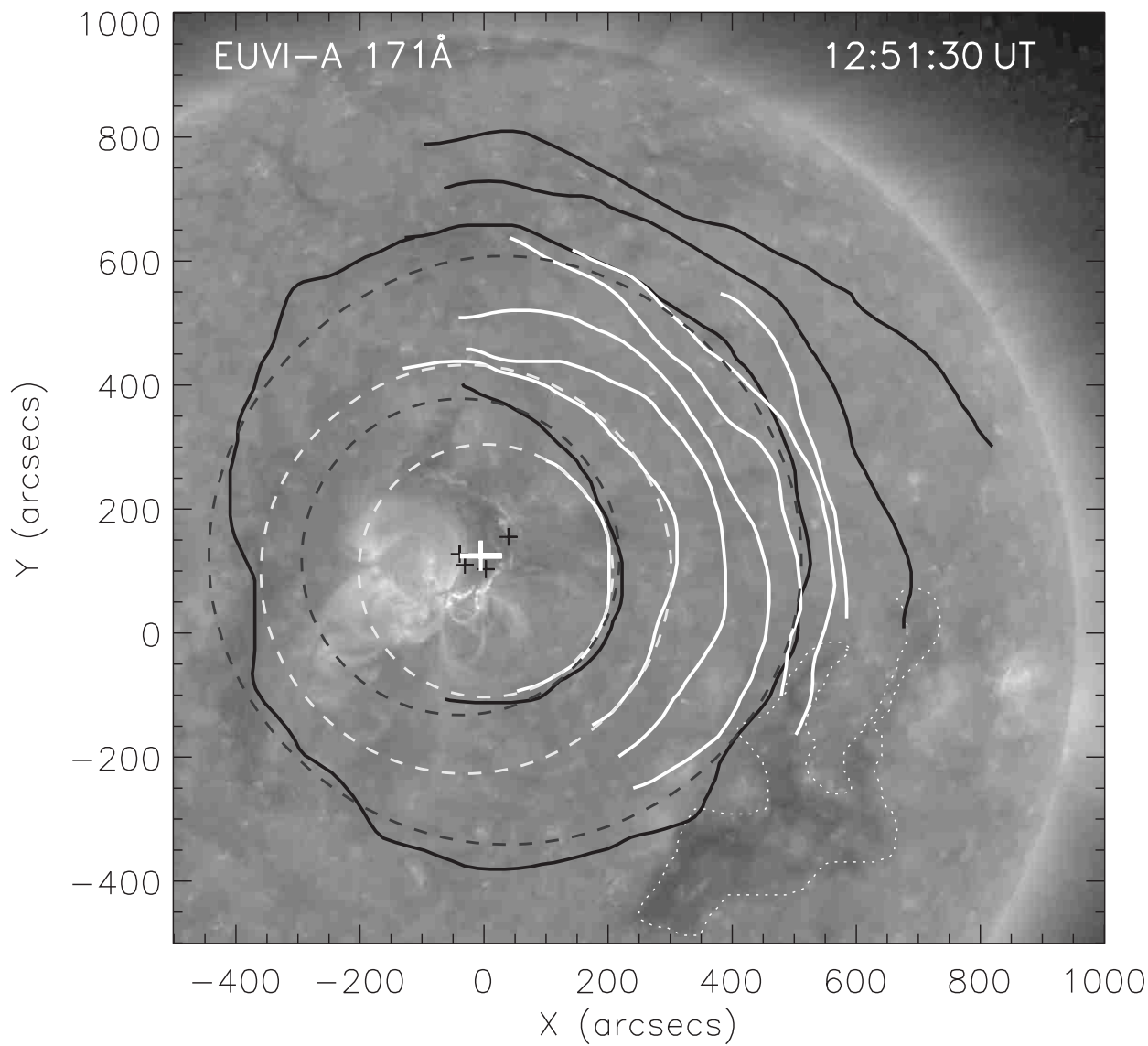}
  \caption{Section of an EUVI 171~{\AA} image. The full lines mark the wave fronts identified in
  171~{\AA} (white) and 195~{\AA} (black) images (cf. Figs.~\ref{euvi_195} and \ref{euvi_171}).
  The dashed lines indicate the circular fits to the two earliest wave fronts observed in 171 and 195~{\AA},
  respectively, which appear as  ellipses in the 2D projected solar image. The wave centers derived
  by this method (black crosses) lie at the NW edge of AR 10956. The white cross marks the mean of the individual wave
  centers derived: $[x_c,y_c]$ = $[-7'',123''] \pm [40'',20'']$. The dotted curve outlines the border of a
  nearby coronal hole, where the wave is refracted and reflected.}
    \label{euvi_image}
\end{figure}

Figures~\ref{euvi_195} and \ref{euvi_171} show running ratio images  (i.e.\ each image is divided by the previous one) 
of the coronal wave observed in the EUVI 195~{\AA} and 171~{\AA} passband, respectively. The coronal wave fronts could be identified and measured in 7~images taken in the 171~{\AA} and 4~images in the 195~{\AA} filter during a period of $\approx$30~min. The wave shows global propagation (see 195~{\AA} image at 13:02~UT) but is most pronounced towards W and NW. The wave ``radiant point" was estimated by applying circular fits to the two earliest
wave fronts observed in 171 and 195~{\AA} after transforming the data from the 2D cartesian $xy$-plane to the 3D
spherical $r\theta \phi$-plane (solar radius, heliographic latitude and longitude). The
center of curvature of the wave was found on the NW border of AR~10956. In
Fig.~\ref{euvi_image}, the determined wave centers together with the strongest wavefront segments
are plotted.
In contrast to Moreton waves, 360$^\circ$ propagation is not uncommon for coronal waves \cite[e.g., discovery event of][]{thompson98}. Simulations by \cite{ofman02} revealed that the wave is strongly refracted/reflected by ARs, 
but part of it can pass through. If the AR is small, the wave can be diffracted into the region behind the AR (passing above and aside), which may be the case in the event under study. We also note that the wave was refracted and reflected at the coronal hole in the SW solar quadrant  (cf.\ Fig.~\ref{euvi_image}). 

Figure~\ref{euvi_kinematics} shows the distance-time diagram of the wave derived by calculating the
mean distance of the wave fronts from the wave center along great circles on the lower corona, 
where the EUVI wave is observed (estimated to be 10~Mm above the 
photosphere). For the kinematics, we used only the strongest wavefront segments (at 13:02~UT the
global wave front) shown in Fig.~\ref{euvi_image}. The linear fit to the distance-time diagram
gives a mean wave velocity of 260~km~s$^{-1}$. 
The quadratic fit yields a start velocity of 460~km~s$^{-1}$, a (constant)
deceleration of $-160$~m~s$^{2}$ and an extrapolated wave launch time (intersect with $x$-axis) of
 12:45~UT. The inset in Fig.~\ref{euvi_kinematics} shows the wavefront velocity
derived by numerical derivative of the measured time-distance data using 3-point linear interpolation,
together with the applied fits. The velocity evolution demonstrates
that the wave decelerates, with the earliest velocities as high as 400--500~km~s$^{-1}$. 
This is considerably faster than the velocities
reported for EIT waves \citep[170--350~km~s$^{-1}$;][]{klassen00}, which we attribute to the much
better cadence of EUVI, allowing us to study the wave's evolution.

Up to now, deceleration of coronal waves was mainly hypothesized from the deceleration observed in
chromospheric Moreton waves recorded with high cadence, and from their combination with (mostly single) EIT wave fronts
lying on the extrapolated kinematical curve of the decelerating Moreton wave \citep{warmuth01,warmuth04a,vrsnak02}.
Evidence for deceleration of coronal waves was provided also from soft X-ray imaging in \cite{warmuth05}. However, in the coronal wave studied in \cite{wills99} the general picture was quite different in that some parts of the wave fronts showed acceleration. Initial acceleration of the wave is actually expected from the theoretical point of view \cite[see Fig.~4 in][]{vrsnak00}.

\subsection{Dynamics of the associated eruptions}

The LASCO/SOHO catalog \citep{yashiro04} reports two CMEs in association with the coronal wave under study:
CME1 was observed at a position angle~PA of 260$^\circ$ in the FOV 3--20~$R_s$ with a mean velocity of 960~km~s$^{-1}$,
CME2 at PA 310$^\circ$ in the FOV 3--10~$R_s$ with 290~km~s$^{-1}$.
Both CMEs were rather poor events, but associated with an interplanetary magnetic cloud
observed in situ at $\approx$1~AU on 21/22 May 2007 by STEREO and Wind \citep{liu08}.
COR1 on STEREO-A did not observe the fast CME1 (which was very faint) but it observed the evolution of CME2 in the low corona. The linear fit gives a mean CME2 velocity of 430~km~s$^{-1}$ in the FOV 1.5--2.5~$R_s$; the quadratic fit
gives a start velocity of 650~km~s$^{-1}$ at 12:48~UT, and a deceleration of $-90$~m~s$^{2}$.

In Fig.~\ref{euvi_kso} we plot a sequence of H$\alpha$ filtergrams which show the earliest sign of the eruption,
two erupting filaments. Their orientation and direction is consistent with the position angles of the two CMEs.
Filament1 (towards W) has disappeared from the H$\alpha$ filter at 12:46 UT, filament2 (towards NW)
at 12:55~UT (see Fig.~\ref{euvi_kso}). The filament evolution appears more complex in EUVI 171~{\AA} images, with fast changes/eruption of the filament system starting between 12:49:00 and 12:51:30 UT.

\section{Discussion and Conclusions}

In Figure~\ref{euvi_summary}, we show a summary plot comprising: a) the distance-time diagram of the coronal wave observed by EUVI/STEREO-A, b) the back-extrapolated (quadratic fit) distance-time diagram of CME1 observed with LASCO/SOHO, c) the distance-time diagram of CME2 observed with COR1/STEREO-A, d) the flare hard X-ray flux recorded by RHESSI, and e) the flare soft X-ray flux recorded by GOES. From the quadratic fit to the EUVI wave kinematics, we estimate the wave's launch time to $\approx$12:45~UT. The real launch may happen somewhat later, since this method 
assumes a point-like origin of the wave. The flare 12--25~keV hard X-ray flux starts rising at 12:50~UT with the first and highest peak at 12:51:30~UT. At this time,  we already observe the first EUVI wave front. Such timing argues against a flare-origin of the wave, since the wave needs time to build up a large amplitude or shock to be observable. 
On the other hand, timing and direction of the erupting filaments indicate that the wave was closely associated with the fast CME1, since filament1 disappeared from the H$\alpha$ filter at 12:46~UT, whereas filament2 remained visible until 12:55~UT. 

\begin{figure}
\epsscale{1}
\plotone{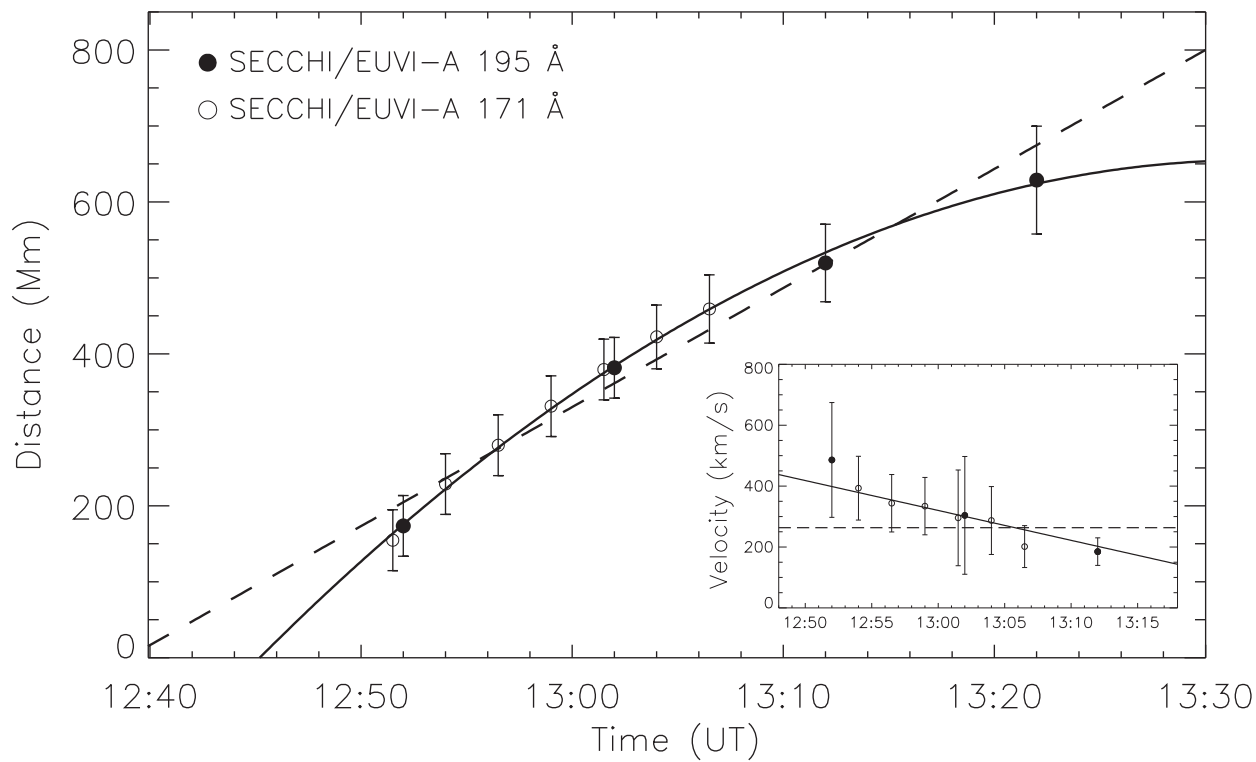}
\caption{Kinematics of the coronal wave observed by EUVI/STEREO-A: Combined distance vs.\ time
diagram derived from the wave fronts observed in 195~{\AA} and 171~{\AA}. The error
bars reflect uncertainties on the calculated wave center as well as on the identification of the wave fronts. The dashed, gray and black lines indicate the linear and quadratic least-squares fits to the time-distance data, respectively. The inset shows the velocity evolution derived by numerical derivative using 3-point linear interpolation (error bars are due to uncertainties in the wave front determination).}
\label{euvi_kinematics}
\end{figure}

\begin{figure}
\epsscale{1}
\plotone{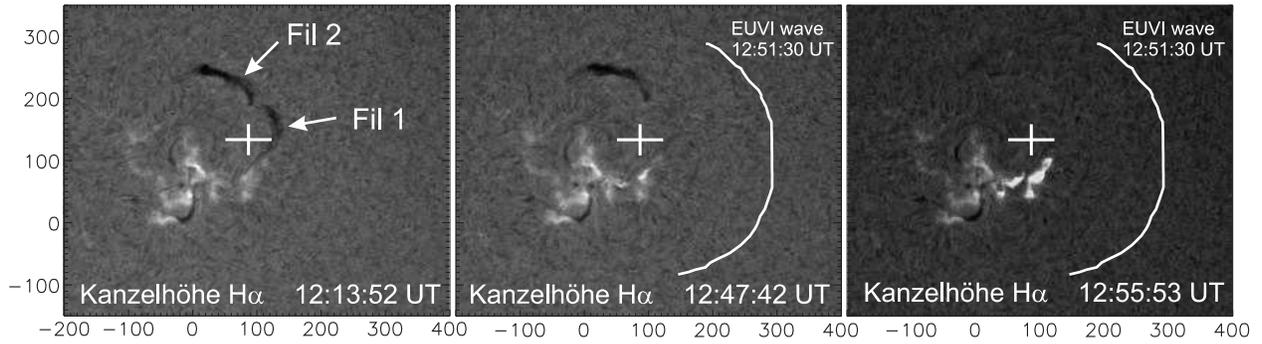}
\caption{Sequence of H$\alpha$ filtergrams recorded at Kanzelh\"ohe Observatory
showing the two erupting filaments. For comparison, we also plot the first EUVI wavefront (12:51:30~UT) and the derived wave ignition center, which have been transformed by $+5.7^\circ$
in longitude and $+0.7^\circ$ in latitude in order to account for the different view from
STEREO-A with respect to Earth. Coordinates are in arcsec.}
\label{euvi_kso}
\end{figure}

\begin{figure}[tbh]
\epsscale{1}
\plotone{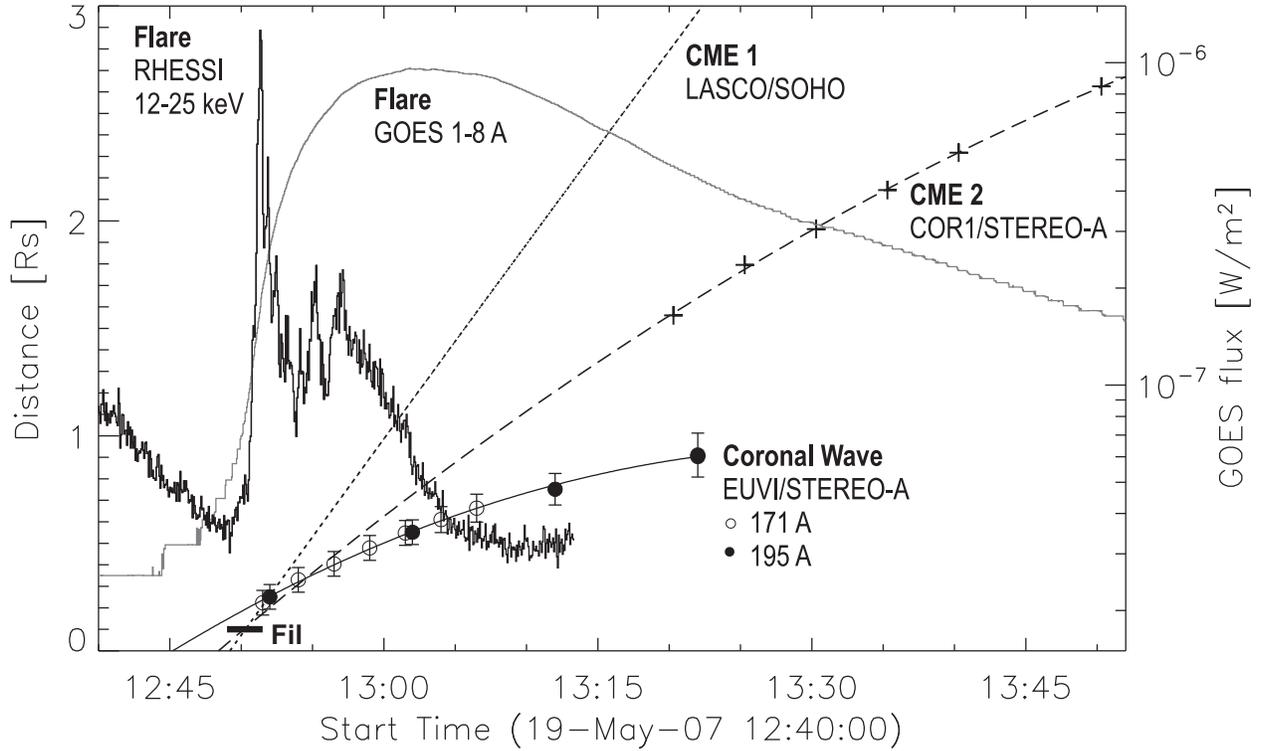}
\caption{Summary plot of the coronal wave kinematics (measured distances: circles; quadratic fit: full lines) together with the flare evolution (GOES 1--8~{\AA} soft X-ray flux: gray curve; RHESSI 12--25 keV hard X-rays: black spiky curve) and kinematics of CME2 observed in COR1/STEREO (pluses; together with quadratic fit) and the back-extrapolated quadratic fit of CME1 observed by LASCO/SOHO (dotted curve). The horizontal bar indicates the start of the fast filament eruption observed by EUVI.}
\label{euvi_summary}
\end{figure}

However, the kinematics of the coronal wave is quite different from the kinematics of the CME's
leading edge  (see Fig.~\ref{euvi_summary}): the wave is slower than both CMEs and 
significantly decelerates, which is a typical characteristics of a large-amplitude MHD simple wave \citep{mann95,vrsnak00}: such a freely propagating
perturbation is powered only temporarily by a source region expansion, which could be due to the
flare-related pressure pulse, due to small scale flare ejecta or due to the CME expanding flanks
(which propagate laterally only over a limited distance). Since perturbation elements with larger
amplitude travel faster than those with smaller amplitude (nonlinearity), the perturbation profile
steepens until finally a discontinuity is formed. As a consequence of energy conservation, the
amplitude of the perturbation decreases with distance, first due to spherical expansion
($\sim$$R^{-2}$), and second, because the crest of the shock travels faster than its trail, causing
broading of the perturbation profile  \citep{landau87}. Consequently, the wave decays into an
ordinary, i.e.\ small amplitude wave propagating with the characteristic speed of the medium. For
waves propagating perpendicular to the magnetic field, this is the magnetosonic speed $v_{\rm ms} =
(v_A^2 + c_s^2)^{1/2}$ with $v_A$ the Alfv\'en velocity and $c_s$ the sound speed. The ``final"
velocity reached by the EUVI wave under study lies in the range $200\pm 50$~km s$^{-1}$ (see
Fig.~\ref{euvi_kinematics}), which is a reasonable value of $v_{\rm ms}$ in the quiet solar corona
\citep[e.g.,][]{mann99}, though we note that there is an ongoing discussion on this subject
\cite[e.g.][]{wang00,wu01,chen02,wills07}.

The observed EUVI wave deceleration together with the closely related timing of the wave and
the erupting filament1/CME1 (in contrast to the flare peak which occurs too late) as well as the
wave front shape which is roughly concentric with filament1, hint at an initiation of the wave by the CME expanding flanks. In such a scenario, the wave is only driven over a limited distance and then decays into an ordinary MHD wave.

\acknowledgements
We thank the SOHO, LASCO, STEREO, EUVI, and RHESSI teams for their open data policy, and the referee
for his/her constructive comments. This work was supported by the Austrian Science Fund (FWF grants P20145-N16 and P20867-N16).

\end{document}